# Laser-Assisted Self-Induced Feshbach resonance for ultracold polar molecule formation

Adrien Devolder,[*] Eliane Luc-Koenig,[*] Osman Atabek,[†] Michèle Desouter-Lecomte,[‡] and Olivier Dulieu[*]

We propose a new type of Feshbach resonance occurring when two different ultracold atoms in their ground state undergo an s-wave collision in the presence of a continuous-wave laser light. The atoms collide in the dissociation continuum of the molecular electronic ground state which is coupled by the light to a rovibrational level of the same electronic ground state: we name this a Laser-Assisted Self-Induced Feshbach Resonance (LASIFR). This mechanism, valid for all polar molecules, is analyzed on the example of ultracold $^{87}$Rb and $^{84}$Sr atoms, for which the laser frequency falls in the THz range. The control of the LASIFR with the laser frequency and intensity allows for a strong increase of the pair probability density at short distances, which tremendously increases the number of atoms pairs transfered toward the absolute ground state level by STImulated Rapid Adiabatic Passage (STIRAP). The LASIFR results in the observation of a standard Fano profile in the pump transition of the STIRAP process, and is also promising for the optical control of the interspecies scattering length without atom losses.

Initially introduced in the context of nuclear reactions [1, 2], and of atomic autoionizing states [3], a Fano-Feshbach resonance (FR) refers to the interaction of a quasibound state of a so-called "closed channel" embedded in the continuum of a so-called "open channel". Anticipated in the early days of evaporative cooling of atomic hydrogen [4], the concept of Magnetic Feshbach Resonance (MFR) has been predicted as a very powerful tool to control the scattering properties of an ultracold atomic gas by tuning the strength of a magnetic field [5]. MFR revolutionized this research field by enabling the long-sought observation of ultracold quantum degenerate gases [6, 7] and the efficient formation of ultracold molecules [8, 9]. The process of magnetoassociation relying on a MFR is at the heart of the successful formation of ultracold heteronuclear alkali-metal diatomic molecules in their absolute ground state, as demonstrated in several spectacular experiments [10–15].

Ultracold polar paramagnetic molecules composed by an alkali-metal atom and an alkaline-earth atom, are promising for instance for quantum simulation of lattice-spin systems [16]. However they have not been created yet in their absolute ground state. It has long been thought that if one of the atoms has no magnetic moment, MFR could not be implemented. A subtle coupling mechanism has been recently invoked to predict MFRs between paramagnetic Rb($^2$S) and non-magnetic Sr($^1$S) atoms [17, 18]. The observation of the very narrow resulting MFR remains however challenging [18]. Another kind of FR that could be implemented for any atomic pair is the Optical Feshbach Resonance (OFR) [19–21]: the dissociation continuum of the molecular electronic ground state associated to the colliding atom pair is coupled to a bound level of an excited electronic state by a laser with a frequency detuned to the red of an atomic transition (Fig.1a, b). An OFR allows for the control of the scattering length with both the laser intensity and frequency. However, the limited radiative lifetime of the excited state induces the formation of vibrationally hot molecules and losses of atoms restraining the efficiency of the scattering length control.

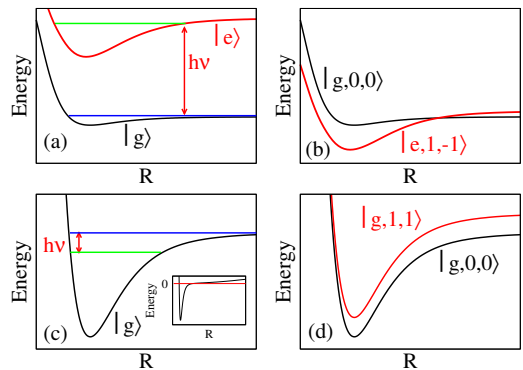

FIG. 1. (Color online) Scheme for an OFR ((a) and (b)), and for the present LASIFR ((c) and (d)). Panels (a) and (c) (resp. (b) and (d)) hold for the undressed (resp. dressed with $n_\gamma$ photons) picture of the potential energy of the ground ($|g\rangle$) and excited ($|e\rangle$) electronic states including the centrifugal barrier $\hbar^2 J(J+1)/(2\mu R^2)$. The dressed ro-electronic states (or channels) are labeled as $|g/e, J, n_\gamma\rangle$. The inset of the panel (c) illustrates the long-range behaviour dominated by the isotropic harmonic trapping potential of the particles (see text).

In this Letter, we propose a new type of FR: the Laser-Assisted Self-Induced Feshbach Resonance (LASIFR), where the closed and open coupled channels are both associated with the ground electronic state $|g\rangle$ of a heteronuclear diatomic molecule

(Fig.1(c), (d)) with an intrinsic permanent electric dipole moment (PEDM) $\mu_{gg}(R)$ varying with the internuclear distance $R$. A continuous-wave laser couples the pair of atoms, colliding at ultracold temperature in s-wave ($J = 0$), to a bound $J = 1$ rovibrational state close to the dissociation limit (Fig.1(c)). This enhances the incident probability density of the atom pair at small $R$. In ongoing experiments, the atoms are often trapped in an optical lattice, occupying a single motional level of the trapping potential. Due to the LASIFR, the adiabatic transfer of the pair population down to the absolute molecular ground state can be strongly increased. Moreover, the LASIFR is suitable for controlling the interspecies scattering length, as with MFR and OFR, thus opening new ways for evaporative cooling for instance. In contrast with OFR, the spontaneous emission during the LASIFR lifetime is negligible, as it involves two levels from the electronic ground-state $|g\rangle$. A similar coupling has been previously proposed in [22] but without establishing the analogy with a FR. Another type of FR induced by radiofrequency field was also proposed in [23].

As for OFR (Fig.1(b)), the resonant pattern becomes obvious in the field-dressed-state picture (Fig.1(d)). The open and closed channels are described by the two dressed ro-electronic states $|g, J = 0, n_\gamma = 0\rangle$ and $|g, J = 1, n_\gamma = 1\rangle$ respectively, where $n_\gamma$ is the number of exchanged photons. The energy difference between the dissociation limit of the two channels equals the photon energy $\hbar\omega_\gamma$, and the channel coupling involving the Rabi frequency $\Omega_\gamma(R)$ is $V_{opt}(R) = \hbar\Omega_\gamma(R) = \frac{1}{\sqrt{3}}\sqrt{\frac{I_\gamma}{2\epsilon_0 c}}\mu_{gg}(R)$, where $I_\gamma$ is the field intensity, $\epsilon_0$ is the vacuum permittivity, $c$ the light velocity and $\frac{1}{\sqrt{3}}$ is the relevant Hönl-London factor. The change of representation from Fig.1(c) to (d) corresponds to the transformation from time-dependent to time-independent Schrödinger equations, thanks to the Floquet theory [24]. We restricted our study to a two-channel case, i.e. to single-photon transitions. This approximation is valid when the maximum $\hbar\Omega_\gamma^{max}$ of the Rabi coupling remains much smaller than the energy difference between the dissociation limits of the channels, or $\Omega_\gamma^{max} \ll \omega_\gamma$.

Here we model the LASIFR to fit the case of a 3D isotropic harmonic optical trap. The properties of the LASIFR in the untrapped case will be discussed in a forthcoming paper. We consider two overlapping Mott insulators with a single atom per site. The atomic trap frequency for the atom X in the optical lattice (OL) at the wavelength $\lambda_{OL}$ and laser intensity $I_{OL}$ is $\omega_X = 2\pi\sqrt{(2\alpha_X I_{OL}/m_X)}/\lambda_{OL}$, where $m_X$ and $\alpha_X$ are the mass and dynamic polarizability at $\lambda_{OL}$. As the masses of Rb and Sr are similar, we consider only the relative motion in the presence of the trapping laser even if a coupling between the relative and the center of mass motions could arise due to the difference in polarizabilities. This leads to the addition in the total Hamiltonian of a harmonic potential $V_{trap} = \frac{1}{2}\mu\omega_R^2 R^2$ with $\omega_R = \sqrt{\frac{m_2\omega_1^2 + m_1\omega_2^2}{m_1 + m_2}}$ and $\mu$ the reduced mass. The trap is characterized by a length $a_\omega = \sqrt{\frac{\hbar}{\mu\omega_R}}$ and prevails at large distance whereas the interactions at short distance are characterized by the scattering length $a$. The dynamics is entirely determined by the ratio $\xi = |a|/a_\omega$. Note that the dependence on $\xi$ of the eigenvalues and eigenfunctions has been determined analytically [25]. We assume that the atom pair initially occupies the lowest trap state with $J = 0$ and energy $E_0^{trap}$ when the LASIFR laser is switched on. The typical temperature that satisfies this situation for Rb and Sr atoms is 1 $\mu$K, leading to a population of 99.3%. The height of the $J = 1$ centrifugal barrier located at $R_b = 172a_0$ is 92.2 $\mu$K.

The introduction of the LASIFR laser induces a coupling between the bound levels of the two dressed but uncoupled potentials, $V_{g,trap}^{0,0}$ and $V_{g,trap}^{1,1}$ where $V_{g,trap}^{J,n_\gamma} = V_g(R) + V_{trap}(R) + \frac{\hbar^2 J(J+1)}{2\mu R^2} + n_\gamma\hbar\omega_\gamma$. The uncoupled vibrational levels are labeled $|g, J, n_\gamma, v\rangle$ while the uncoupled trap motional states $|g, J, n_\gamma, N\rangle$. The intensity of the laser leading to the LASIFR is ramped adiabatically from zero to $I_\gamma$ such that the population of the atom pair occupies a well defined discrete eigenstate $|\Psi_i\rangle$ of the full Hamiltonian. The eigenfunction $\Psi_i(R)$, expressed in the R-representation, are two component functions $\psi_{g00}^i(R)$ and $\psi_{g11}^i(R)$ solutions of the coupled system :

$$[\hat{T} + V_{g,trap}^{J=0,n_\gamma=0}(R) - E_i]\psi_{g00}^i(R) = -\hbar\Omega_\gamma(R)\psi_{g11}^i(R)$$
$$[\hat{T} + V_{g,trap}^{J=1,n_\gamma=1}(R) - E_i]\psi_{g11}^i(R) = -\hbar\Omega_\gamma(R)\psi_{g00}^i(R)$$
(1)

where $\hat{T} = -\frac{\hbar^2}{2\mu}\frac{d^2}{dR^2}$.

We investigate here the possibility to create ultracold polar $^{87}$Rb$^{84}$Sr molecules using LASIFR. The choice of isotopes is determined by the favorable values of the interspecies scattering length for mixing Bose-Einstein condensates [26]. We rely on the accurate molecular data already used in our previous work [27]. The atomic frequencies in an optical lattice at $\lambda_{OL} = 1064$ nm are $\omega_{Rb} = 2\pi \times 122$ kHz and $\omega_{Sr} = 2\pi \times 72$ kHz, giving $\omega_R = 2\pi \times 100$ kHz. The relevant parameters of the LASIFR for $^{87}$Rb$^{84}$Sr are listed in Table 1 of the Supplementary Material. As the resonant state, we consider the $v_{res} = 56$ level of the uncoupled $X^2\Sigma^+$ electronic ground state, with binding energy $E_{v_{res}} = 6.16$ cm$^{-1}$ (0.18 THz). It corresponds to the largest value of the squared PEDM matrix element $|\langle g, J = 1, n_\gamma = 1, v| \mu_{gg} |g, J = 0, n_\gamma = 0, N = 0\rangle|^2$ between the uncoupled $X^2\Sigma^+$ rovibrational levels and the uncoupled lowest motional state of the trap (See Fig. 1 of the Supplementary Material). The broad shape of the distribution suggests that the choice of the resonant state is flexible.

The equations (1) are solved with the Mapped Fourier Grid Hamiltonian (MFGH) method [28, 29], using a grid extending from $R_{min} = 5a_0$ to $R_{max} = 5000a_0$, covering the spatial extension of the lowest trap state $|g, J = 0, n_\gamma = 0, N = 0\rangle$. Calculations are performed for different values of intensities $I_\gamma$ and detunings $\Delta_\gamma$ and the eigenstates $|\Psi_i\rangle$ are numbered in ascending order of energy. The detuning is defined as $\Delta_\gamma = (E_{v_{res}} + \hbar\omega_\gamma) - E_0^{\text{trap}}$. The variation of the eigenenergies with $\Delta_\gamma$ presented in Fig.2 of the Supplementary Material, show that the occupied coupled eigenstate is $|\Psi_{124}\rangle$ (resp. $|\Psi_{123}\rangle$ ) for negative (resp. positive) detuning. Fig.2 displays the probability density $|\Psi_{123}(R)|^2$ for various $I_\gamma$ and $\Delta_\gamma$ values. At short distances, the component $\psi^i_{g11}(R)$ on the close channel dramatically increases with increasing $I_\gamma$ (at fixed $\Delta_\gamma = 0.01$ MHz, Fig.2(a)) and decreasing $\Delta_\gamma$ (at fixed $I_\gamma = 100$ W/cm$^2$, Fig.2(b)). A similar pattern close to a MFR has been previously reported [30]. However in the case of a LASIFR, the two varying parameters offer flexible control opportunities: the tuning of the field frequency around the resonance condition through $\Delta_\gamma$ is balanced by $I_\gamma$ which determines the coupling between the resonant state and the trap state.

The increase of the probability density at short distance is related to the variation of the scattering length $a$. While the atoms collide inside a trap, it can be determined in the intermediate region $R_{vdW} \ll R \ll a_\omega$ (where $R_{vdW}$ is the van der Waals length) where the wavefunctions $\Psi_{124}(R)$ and $\Psi_{123}(R)$ are proportional to $(R-a)$ (see Supplementary Material). The variation of $a$ with $\Delta_\gamma$ and $I_\gamma$ is displayed on Fig.3. It is similar to the one obtained from for the genuine scattering state involved in MFR. A fit of the scattering length to the analytical formula $a = a_{bg}\left(1 - \frac{\Gamma}{\Delta_\gamma}\right)$ ($a_{bg}$ being the background scattering length i.e. the scattering length in the absence of LASIFR laser) allows for the determination of the resonance width $\Gamma$ which varies linearly with $I_\gamma$, while being constant in the MFR case. For instance we obtain $\Gamma = 2\pi(1.45 \times 10^{-4})$ MHz at $I_\gamma = 1000$ W/cm$^2$, and $\Gamma = 2\pi(6.88 \times 10^{-4}$ MHz

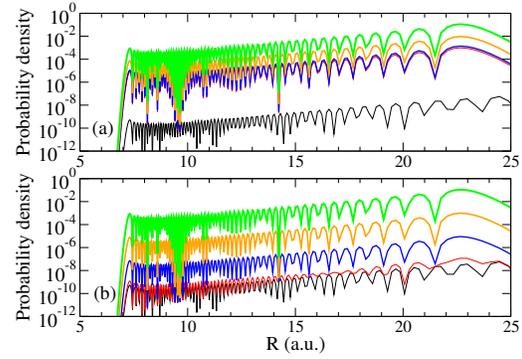

FIG. 2. (Color online) Total probability density of the eigenstate $|\Psi_{123}\rangle$ at short distances. (a) At $\Delta_\gamma = 0.01$ MHz and for different intensities $I_\gamma$=0 W/cm$^2$ (black), 100 W/cm$^2$ (red), 500 W/cm$^2$ (blue), 1000 W/cm$^2$ (orange). (b) At $I_\gamma = 100$ W/cm$^2$ and for different detunings $\Delta_\gamma$=10 MHz (black), 1 MHz (red), 0.1 MHz (blue), 0.01 MHz (orange). The probability density of the uncoupled state $|g, 1, 1, v = 56\rangle$ is drawn in green.

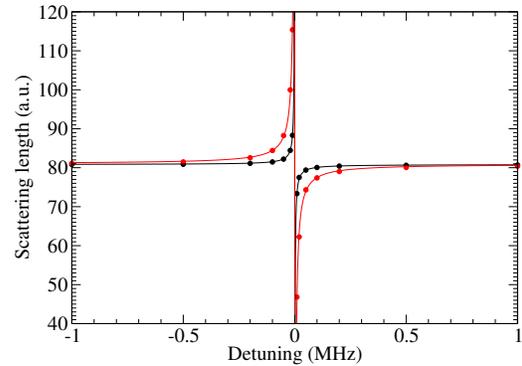

FIG. 3. (Color online) Variation with $\Delta_\gamma$ of the scattering length $a$ for a ($^{87}$Rb,$^{84}$Sr) pair in a 3D isotropic harmonic trap with frequency $\omega_R = 2\pi \times 100$ kHz, for two different intensities $I_\gamma$ =1000 W/cm$^2$ (black) and $I_\gamma$ =5000 W/cm$^2$ (red) of the LASIFR laser. The circles are the computed values and the lines are their analytical fit (see text).

at $I_\gamma = 5000$ W/cm$^2$, which are rather important intensities.

The enhancement of the probability density at short range is also promising for improving schemes aiming at creating ultracold molecules in their absolute ground state using STIRAP [31]. Three lasers would then be involved: a cw laser for the LASIFR, and two optical pulsed lasers for the STIRAP relying on an excited electronic state $|e'\rangle$. In [27], we demonstrated that the population transfer of ultracold RbSr molecules from the lowest trap state toward the $(v'' = 0, J'' = 0)$ level of the $X^2\Sigma^+$ ground state was rather inefficient, due to the poor spatial overlap of the wave functions of the trap state $|g, 0, 0, N = 0\rangle$ and of the deeply-bound vibrational levels $|e', v'\rangle$. In the presence of a LASIFR, this overlap is greatly enhanced. We report in Fig. 4 the huge variation of the squared transition dipole moment (TDM) between the LASIFR dressed state $|\Psi_i\rangle$ ($i = 123, 124$), and $|e' = (2)\Omega = 1/2, v' = 3\rangle$, as a function of $\Delta_\gamma/\sqrt{(I_\gamma)}$ for four different intensities $I_\gamma$. A Fano profile [3], induced by the interference between the contribution of the two radial components $\psi^i_{g,1,1}$ and $\psi^i_{g,0,0}$ involved in the transition toward $|(2)\Omega = 1/2, v' = 3\rangle$, characterizes this variation. The values of the calculated asymmetry parameter $q$ of the Fano profile are given in Table 3 of the Supplementary Material. The possibility of tuning $q$ with $I_\gamma$ is another advantage of LASIFR upon MFR. The $1/\sqrt{I_\gamma}$ dependence of $q$ induces a $\sqrt{I\gamma}$ variation for the detuning of the minimum $\Delta_{min}$, leading to the invariance of $\Delta_{min}/\sqrt{(I_\gamma)}$. The high $q$ values imply that the detuning of the maximum $\Delta_{max}$ stays close to zero for the considered intensities. This explains the universal behavior observed in Fig. 4. Furthermore, the high $q$ value also explains why the difference between $\Delta_{min}$ and $\Delta_{max}$ is much larger than $\Gamma$ (see Supplementary Material for the expressions of $q$, $\Delta_{min}$, $\Delta_{max}$ and $\Gamma$).

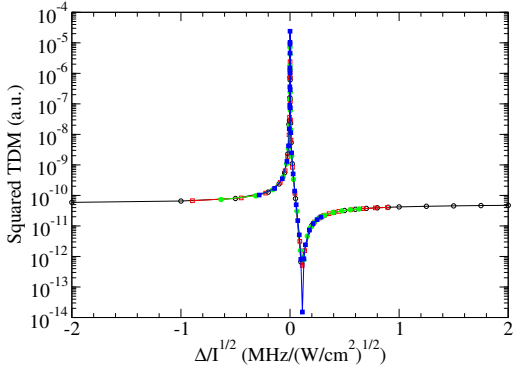

FIG. 4. (Color online) RbSr squared transition dipole moments (squared TDM) between the LASIFR dressed state $|\Psi_{124}\rangle$ for negative detuning or $|\Psi_{123}\rangle$ for positive detuning, and the vibrational level $|e' = (2)\Omega = 1/2, v' = 3\rangle$ of the $(2)\Omega = 1/2$ excited molecular state (correlated to the $Rb(5p\,^2P_{1/2})+Sr(5s^2\,^1S)$ asymptote) versus $\Delta_\gamma/\sqrt{I_\gamma}$, for $I_\gamma$=100 W/cm$^2$ (black line with open circles), 500 W/cm$^2$ (red line with open squares), 1000 W/cm$^2$ (green line with closed circles) and 5000 W/cm$^2$ (blue line with closed squares).

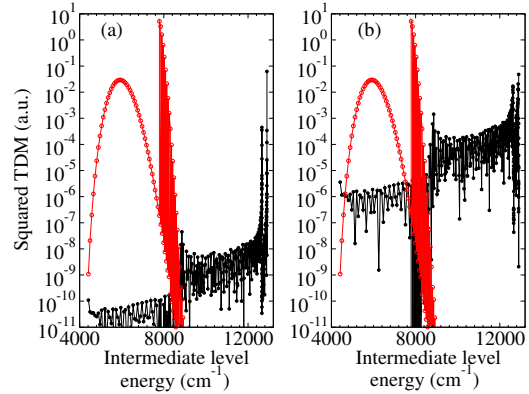

FIG. 5. (Color online) RbSr squared transition dipole moment (squared TDM) for the pump (black lines) and dump (red lines) transitions relevant for the STIRAP population transfer from the eigenstate $|\Psi_{124}\rangle$ to the absolute ground state level $v'' = 0$, as a function of the excitation energy of the intermediate level $|(2)\Omega = 1/2, v'\rangle$. (a) STIRAP far from a LASIFR; (b) close to a LASIFR with $I_\gamma = 500$ W/cm$^2$ and $\Delta_\gamma = -0.01$ MHz.

The resulting enhancement of the STIRAP transfer efficiency is illustrated in Fig.5. The choice of the optimal intermediate level $|e', v'\rangle$ is dictated by two requirements, assuming equal laser intensities for both pump and dump STIRAP transitions: the corresponding squared TDMs must be of the same order of magnitude, and must be sufficiently large (typically larger than $10^{-6}$ a.u., see for instance [32]). Off resonance (Fig5(a), or close to resonance (Fig. 5 (b)), the pump and dump TDMs have equal values for an intermediate level with about 7500 cm$^{-1}$ of excitation energy from the lowest trap state (energy of the pump transition). In the presence of LASIFR, the squared TDMs of the pump transition increase by 5 orders of magnitude as compared to their values in the off-resonant situation. Furthermore, another crossing is located around 4600 cm$^{-1}$, corresponding to $|(2)\Omega = 1/2, v' = 3\rangle$, giving more flexibility for the choice of the intermediate level. The main characteristics of these transitions are reported in Table 2 of the Supplementary Material.

In this work we have shown the implementation of a new laser induced and controlled mechanism, namely LASIFR. The specific Feshbach resonance in play for this mechanism is induced by a narrow bandwidth laser source, which generates a radiative coupling within the electronic molecular ground state with a $R$-dependent permanent electric dipole moment. The proposed mechanism is promising for the creation of ultracold ground-state molecules which hardly exhibit magnetically-tunable Feshbach resonances, as well as for the control of the scattering properties of non-identical atoms. The coupling can be controlled by both the laser intensity and its frequency detuning. According to the literature [33], suitable sources in the THz frequency domain should be available in the case of $^{87}$Rb and $^{84}$Sr ultracold atom pairs. For the formation of molecules, as illustrated in this Letter, the LASIFR could indirectly participate by facilitating a STIRAP process. A direct formation of molecules by adiabatic passage using a LASIFR induced by a pulsed laser is an envisioned option.

### ACKNOWLEDGMENTS

The authors acknowledge the support of the ANR BLUESHIELD (Grant No. ANR-14-CE34-0006), the GDR THEMS and the use of the computing center MésoLUM of the LUMAT research federation (FR LUMAT 2764).

* Laboratoire Aimé Cotton, CNRS, Université Paris-Sud, ENS Paris-Saclay, Université Paris-Saclay, Bât. 505, 91405 Orsay Cedex, France
† Institut des Sciences Moléculaires d'Orsay, CNRS, Université Paris-Sud, Université Paris-Saclay, Bât. 520, 91405, Orsay, France
‡ Laboratoire de Chimie Physique, CNRS, Université Paris-Sud, Université Paris-Saclay, Bât. 349, 91405, Orsay France

[1] H. Feshbach, Ann. of Phys. **5**, 357 (1958).
[2] H. Feshbach, Ann. of Phys. **19**, 287 (1962).
[3] U. Fano, Phys. Rev. **124**, 1866 (1961).
[4] W. C. Stwalley, Phys. Rev. Lett. **37**, 1628 (1976).
[5] E. Tiesinga, B. J. Verhaar, and H. T. C. Stoof, Phys. Rev. A **47**, 4114 (1993).
[6] E. Timmermans, P. Tommasini, M. Hussein, and A. Kerman, Phys. Rep. **315**, 199 (1999).
[7] C. Chin, R. Grimm, P. Julienne, and E. Tiesinga, Rev. Mod. Phys. **82**, 1225 (2010).
[8] L. D. Carr, D. DeMille, R. V. Krems, and J. Ye, New J. Phys. **11**, 055049 (2009).
[9] O. Dulieu and C. Gabbanini, Rep. Prog. Phys. **72**, 086401 (2009).
[10] K.-K. Ni, S. Ospelkaus, M. H. G. de Miranda, A. Pe'er, B. Neyenhuis, J. J. Zirbel, S. Kotochigova, P. S. Julienne, D. S. Jin, and J. Ye, Science **322**, 231 (2008).
[11] T. Takekoshi, L. Reichsöllner, A. Schindewolf, J. M. Hutson, C. R. LeSueur, O. Dulieu, F. Ferlaino, R. Grimm, and H.-C. Nägerl, Phys. Rev. Lett. **113**, 205301 (2014).
[12] J. W. Park, S. A. Will, and M. W. Zwierlein, Phys. Rev. Lett. **114**, 205302 (2015).
[13] M. Guo, B. Zhu, B. Lu, X. Ye, F. Wang, R. Vexiau, N. Bouloufa-Maafa, G. Quéméner, O. Dulieu, and D. Wang, Phys. Rev. Lett. **116**, 205303 (2016).
[14] T. M. Rvachov, H. Son, A. T. Sommer, S. Ebadi, J. J. Park, M. W. Zwierlein, W. Ketterle, and A. O. Jamision, Phys. Rev. Lett. **119**, 143001 (2017).
[15] F. Seeßelberg, N. Buchheim, Z.-K. Lu, T. Schneider, X.-Y. Luo, E. Tiemann, I. Bloch, and C. Gohle, Phys. Rev. A **97**, 013405 (2018).
[16] A. Micheli, G. K. Brennen, and P. Zoller, Nature Physics **2**, 341 (2006).
[17] P. S. Żuchowski, J. Aldegunde, and J. M. Hutson, Phys. Rev. Lett. **105**, 153201 (2010).
[18] V. Barbé, A. Ciamei, B. Pasquiou, L. Reichsöllner, F. Schreck, P. S. Żuchowski, and J. M. Hutson, Nat. Phys. **14**, 881 (2018).
[19] P. O. Fedichev, Y. Kagan, G. V. Shlyapnikov, and J. T. M. Walraven, Phys. Rev. Lett. **77**, 2913 (1996).
[20] F. K. Fatemi, K. M. Jones, and P. D. Lett, Phys. Rev. Lett. **85**, 4462 (2000).
[21] T. L. Nicholson, S. Blatt, B. J. Bloom, J. R. Williams, J. W. Thomsen, and J. Ye, Phys. Rev. A **92**, 022709 (2015).
[22] S. Kotochigova, Phys. Rev. Lett. **99**, 073003 (2007).
[23] Y. Ding, J. P. D'Incao, and C. H. Greene, Phys. Rev. A **95**, 022709 (2017).
[24] J. H. Shirley, Phys. Rev. **138**, B979 (1965).
[25] T. Busch, B.-G. Englert, K. RzaŻewski, and M. Wilkens, Fond. Phys. **28**, 549 (1998).
[26] A. Ciamei, J. Szczepkowski, A. Bayerle, V. Barbé, L. Reichsöllner, S. M. Tzanova, C.-C. Chen, B. Pasquiou, A. Grochola, P. Kowalczyk, W. Jastrzebski, and F. Schreck, Phys. Chem. Chem. Phys. **20**, 26221 (2018).
[27] A. Devolder, E. Luc-Koenig, O. Atabek, M. Desouter-Lecomte, and O. Dulieu, Phys. Rev. A **98**, 053411 (2018).
[28] V. Kokoouline, O. Dulieu, R. Kosloff, and F. Masnou-Seeuws, J. Chem. Phys **20**, 9865 (1999).
[29] K. Wilner, O. Dulieu, and F. Masnou-Seeuws, J. Chem. Phys. **120**, 548 (2004).
[30] P. Pellegrini, M. Gacesa, and R. Côté, Phys. Rev. Lett. **101**, 053201 (2008).
[31] U. Gaubatz, P. Rudecki, S. Schiemann, and K. Bergmann, J. Chem. Phys. **92**, 5363 (1990).
[32] D. Borsalino, B. Londono-Florez, R. Vexiau, O. Dulieu, N. Bouloufa-Maafa, and E. Luc-Koenig, Phys. Rev. A **90**, 033413 (2014).
[33] D. J. Paul, Laser Photonics Rev. **4**, 610 (2010).

# Supplementary Material

## Molecular parameters relevant for a LASIFR

In table 1, we list the molecular parameters relevant for the implementation of a LASIFR for the $^{87}$Rb$^{84}$Sr molecule, and compare them to those of a typical heteronuclear alkali-metal dimer, namely $^{39}$K$^{87}$Rb. The reported PEDM matrix element corresponds to $\langle \mu_{gg}(v_{res}; N = 0) \rangle = \langle g, J = 1, n_\gamma = 1, v_{res} | \mu_{gg} | g, J = 0, n_\gamma = 0, N = 0 \rangle$, discussed in the main text.

Most numbers are of the same order of magnitude, supporting the transposable character of LASIFR that could be implemented for a large class of diatomic polar molecules. This is also examplified in Fig. 1, where the PEDM matrix elements $\langle \mu_{gg}(v; N = 0) \rangle$ are displayed for various ground-state vibrational levels $v$. The pattern is similar for both $^{87}$Rb$^{84}$Sr and $^{39}$K$^{87}$Rb: the level $v_{res}$ is taken at the main maximum, located for both species in the same frequency range for the requested laser.

The variation of the eigenenergies of the states resulting from the LASIFR with the detuning $\Delta_\gamma$ for fixed $I_\gamma = 100$ W/cm$^2$ is represented in Fig. 2. The occupied coupled eigenstate adiabatically correlated to the $(N = 0, J = 0)$ field free trap level is $|\Psi_{124}\rangle$ for negative detuning or $|\Psi_{123}\rangle$ for positive detuning.

|  | $^{87}$Rb$^{84}$Sr | $^{39}$K$^{87}$Rb |
|---|---|---|
| Background scattering length $a_{bg}$ (in $a_0$) | 80 | 876 |
| Harmonic trap characteristic length $a_\omega$ (in $a_0$) | 919 | 1158 |
| van der Waals distance $R_{vdw}$ (in $a_0$) | 154.9 | 141.7 |
| Vibrational quantum number of the resonant state $v_{res}$ | 56 | 89 |
| Energy of the optimal resonant state $E_{v_{res}}$ (in cm$^{-1}$) | -6.2 | -14.6 |
| Inner turning point of the resonant state $R_{in}^{res}$ (in $a_0$) | 7.3 | 5.7 |
| Outer turning point of the resonant state $R_{out}^{res}$ (in $a_0$) | 22.8 | 21.1 |
| Energy of the first trap state J=0 $E_0^{trap}$ (in cm$^{-1}$) | $5.3 \times 10^{-6}$ | $7.41 \times 10^{-6}$ |
| Inner turning point of the first trap state $R_{in}^{trap}$ (in $a_0$) | 7.2 | 5.6 |
| Outer turning point of the first trap state $R_{out}^{trap}$ (in $a_0$) | 1645 | 2440 |
| PEDM matrix element (in a.u.) | $2.0 \times 10^{-5}$ | $4.0 \times 10^{-5}$ |

Table 1: Main parameters characterizing the molecular system

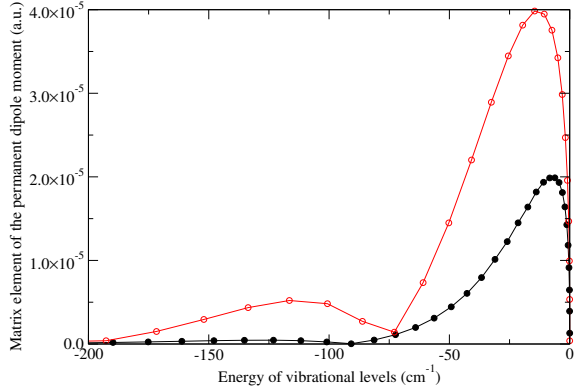

Figure 1: (Color online) Matrix elements of the permanent dipole moment $|\langle g,1,1,v|\mu_{gg}|g,0,0,N=0\rangle|$ between the uncoupled first $(N=0, J=0)$ trap state and the uncoupled ro-vibrational states $(v, J=0)$ of the $X^2\Sigma^+$ electronic ground state of $^{87}$Rb$^{84}$Sr (black) and $^{39}$K$^{87}$Rb (red) as a function of the vibrational energy $E_v$.

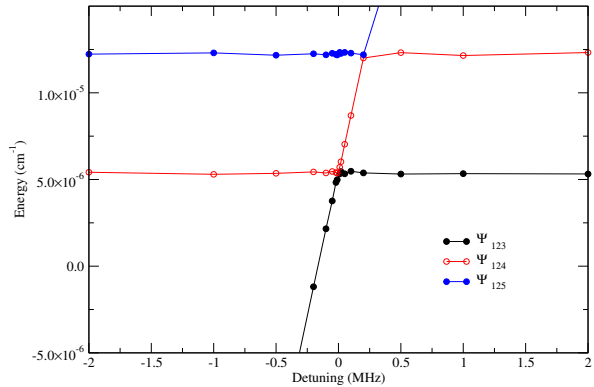

Figure 2: (Color online) Variation of the energy of the first two J=0 trap states (horizontal lines) perturbed by the $(v_{res}, J=1)$ rovibrational state of the $X^2\Sigma^+$ molecular potential (oblique line) in the presence of the LASIFR laser with intensity $I_\gamma = 100$ W/cm$^2$ in $^{87}$Rb$^{84}$Sr, with the detuning $\Delta_\gamma$ of the laser. Due to the interaction, the energy curves exhibit avoided crossings, where there is a change in the numbering of the trap states. This occurs for vanishing $\Delta_\gamma$ for the lowest trap state and at small positive energy for excited trap state.

## Determination of the scattering length for the field-dressed ($^{87}$Rb,$^{84}$Sr) pair

As discussed in the text, we consider, as the initial state of the atom pair, the lowest motional level of the trapping potential, which thus is not a real scattering state. Nevertheless, the scattering length can be determined by locating the distance where the extrapolation in the linear region (between 200 $a_0$ and 400 $a_0$) of the radial component $\psi_{g,0,0}(R)$ vanishes (Fig.3). In this R-range the $X^2\Sigma^+$ potential decreasing asymptotically proportionnaly to $-1/R^6$ becomes negligible meanwhile the trapping potential increasing as $R^2$ is not yet significant so that the $J=0$ near-threshold wavefunction behaves has a free spherical wave with a linear variation. According to Table 1, this assumption is reasonable for the ($^{87}$Rb,$^{84}$Sr) pair, as the ratio $\xi = \frac{a_{bg}}{a_\omega}$ is rather small. This scattering length characterizes the contribution of the short-range interactions, in particular those due to the molecular potential and to the coupling $\Omega(R)$ due to the LASIFR laser. This coupling is of short-range type due to the very different spatial localisation of the probability densities of the bound $J = 1$ resonant state.

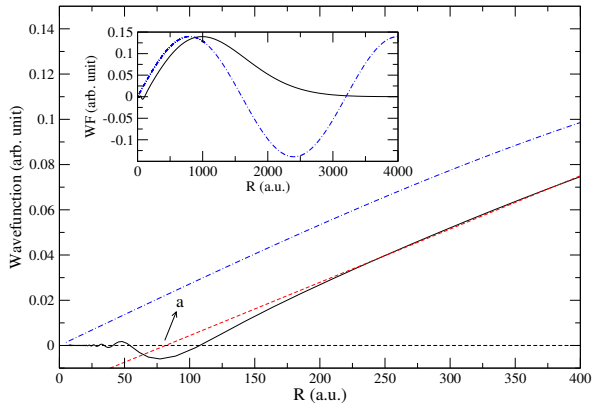

Figure 3: (Color online) Determination of the scattering length of a trap state for the ($^{87}$Rb,$^{84}$Sr) pair. The inset shows the full wave function (black line) decaying at the edge of the trap. Red dashed line: linear extrapolation of the trap state wave function. Blue dot-dashed line: wave function of a free scattering state at the same energy $E_0^{trap}$ than the trap state.

## Characterization of the LASIFR with a Fano profile

The so-called "Fano profile" is the asymmetric shape of the strength of a transition between a discrete state and a quasi-discrete one, i.e. a discrete state coupled to a continuum. Then quantum interferences occur between the two different excitation paths, the direct transition between the two bound wavefunction and the indirect one through the continuum part of the quasi-discrete state wave function.. This is the case of the pump transition of the STIRAP scheme invoked in the paper for $^{87}$Rb$^{84}$Sr. We first report in Table 2 the main characteristics of the two optimal intermediate levels for STIRAP transitions identified in our study (see Figure 5 of the main paper).

|  | First STIRAP | Second STIRAP |
|---|---|---|
| $v_i$ | $\|\Psi_i\rangle$ | $\|\Psi_i\rangle$ |
| $E_i$ (cm$^{-1}$) | 5.43 (-6) | 5.43 (-6) |
| $v''$ | 0 | 0 |
| $E''$ (cm$^{-1}$) | 1054.3406 | 1054.3406 |
| $v'$ | 3 | 43 |
| $E'$ (cm$^{-1}$) | 7926.4524 | 5071.5798 |
| $E_{\text{pump}}$ (cm$^{-1}$) | 4573.5 | 7507.4 |
| $\|d_{ie}\|^2$ (a.u.) | 1.05 (-6) | 3.11 (-6) |
| $E_{\text{dump}}$ (cm$^{-1}$) | 5627.9 | 8561.7 |
| $\|d_{ef}\|^2$ (a.u.) | 1.27 (-6) | 1.79 (-6) |

Table 2: Characteristics of the optimal transition for the two STIRAP schemes relying on an intermediate level belonging to the state $(2)\Omega = \frac{1}{2}$ in the presence of a LASIFR. The initial level is the J=0 lowest trap state and and the final level is the rovibrational ground state ($v'' = 0, J = 0$). The vibrational quantum number $v'$ and the binding energy $E'$ are given for the intermediate level. The energies $E_{\text{pump}}$ and $E_{\text{dump}}$ and the related squared transition dipole moments $|d_{ie}|^2$ and $|d_{ef}|^2$ of the pump and dump transitions are also reported. Number in parenthesis hold for powers of 10. The index $i$ of the initial state holds for $|\Psi_i\rangle$ with $i = 123$ or $124$.

Taking the example of the first STIRAP scheme above, the corresponding squared transition matrix element is given by

$$|\langle \Psi_i | \mu_{ge} | e, v' = 3 \rangle|^2 = |\langle g, 0, 0, N = 0 | \mu_{ge} | e, v' = 3 \rangle|^2 \frac{|q+\epsilon|^2}{1+\epsilon^2} \qquad (1)$$

where i=123 or 124, $\mu_{ge}(R)$ is the transition dipole moment between the X$^2\Sigma^+$ ground state and the $(2)\Omega = \frac{1}{2}$ excited electronic state. The dimensionless asymmetric $q$ parameter is defined as

$$q = \frac{\hbar \omega_R}{\pi} \sqrt{\frac{8\hbar c \epsilon_0}{I}} \frac{\langle g, 1, 1, v_{res} | \mu_{ge} | e, v' = 3 \rangle}{\langle g, 1, 1, v_{res} | \mu_{gg} | g, 0, 0, N = 0 \rangle \langle g, 0, 0, N = 0 | \mu_{ge} | e, v' = 3 \rangle}, \qquad (2)$$

$\epsilon$ is the ratio of the detuning of the laser addressing the transition, and the strength of the coupling matrix element inducing the interference

$$\epsilon = \frac{\Delta_\gamma}{\Gamma/2}, \qquad (3)$$

with

$$\Gamma = \frac{\pi I}{6\hbar \omega_R c \epsilon_0} |\langle g, 1, 1, v_{res} | \mu_{gg} | g, 0, 0, N = 0 \rangle|^2. \qquad (4)$$

The values of the $q$ parameter for various different intensities calculated with eq. (2) are listed in Table 3, revealing their high magnitude for the considered transition in $^{87}$Rb$^{84}$Sr. The used values for the matrix elements are : $\langle g, 1, 1, v_{res} | \mu_{gg} | g, 0, 0, N = 0 \rangle = 2.0 \times 10^{-5}$ a.u., $\langle g, 0, 0, N = 0 | \mu_{ge} | e, v' = 3 \rangle = 7.8 \times 10^{-6}$ a.u. and $\langle g, 1, 1, v_{res} | \mu_{ge} | e, v' = 3 \rangle = 9.1 \times 10^{-3}$ a.u. The detunings of highest and smallest values of the profile are given by $\Delta_{min} = -\Gamma q/2 \propto \sqrt{I}$, and $\Delta_{max} = \Gamma/(2q) \propto I^{3/2}$.

| $I$ (W/cm$^2$) | $q$ |
|---|---|
| 100 | -99645 |
| 500 | -44562 |
| 1000 | -31510 |
| 5000 | -14092 |

Table 3: Asymmetry parameter $q$ of the Fano profile for different intensities